\documentclass[a4paper,11pt]{article}

\usepackage{amsmath}
\usepackage{amsfonts}
\usepackage{amssymb}
\usepackage{pstcol}
\usepackage{multido}
\usepackage[all]{xy}

\definecolor{light-blue}{rgb}{0.8,0.85,1}
\definecolor{light-red}{rgb}{1,0.8,0.8}
\newcommand{\Show}[2]{\psshadowbox[fillstyle=solid,fillcolor=#1]{\txt{#2}}}
\newcommand{\ShowX}[3]{{\Show{#1}{\begin{minipage}[t]{#2}\center{#3}\end{minipage}}}}

\begin{document}

\title{Principles of Equivalence: Their Role in Gravitation Physics and Experiments that Test Them\footnote{Appeared in C. L\"ammerzahl, C.W.F. Everitt, F.W. Hehl (eds.): {\it Gyrod, Clocks, Interferometers...: Testing Relativistic Gravity in Space}, Springer--Velag 2001}} 

\author{Mark~P.~Haugan$^1$ and C. L\"ammerzahl$^2$ \\
$^1$ Purdue University, West Lafayette, IN 47907, USA\\
$^2$ Department of Physics, University of Konstanz, Fach M674, 78457 Konstanz, Germany}

\maketitle   

\begin{abstract}
Modern formulations of equivalence principles provide the
foundation for an efficient approach to understanding and
organizing the structural features of gravitation field theories.
Since theories' predictions reflect differences in their structures,
principles of equivalence also support an efficient experimental
strategy for testing gravitation theories and for exploring the
range of conceivable gravitation physics.  These principles focus
attention squarely on empirical consequences of the fundamental
structural differences that distinguish one gravitation theory from
another.  Interestingly, the variety of such consequences makes it
possible to design and perform experiments that test equivalence
principles stringently but do so in markedly different ways than
the most familiar experimental tests.
\end{abstract}

\section{Equivalence Principles and the Structure of Gravitation Theories}

\subsection{From the weak to Einstein's Equivalence principle}

Since the time of the Renaissance observations indicating
that bodies fall in a gravitational field in a way that is
independent of their internal composition and structure have
been considered remarkable.  Clearly Newton thought so since
he deemed it necessary to perform pendulum experiments to
verify this property of freefall as precisely as he could
before he published laws of motion and universal gravitation
that predict it \cite{Newton}.  Einstein also found this
property of freefall remarkable.  The insight he gained
by reflecting on it in his famous elevator \emph{Gedanken}
experiment \cite{Einstein} is communicated by what we
now call the Einstein equivalence principle (EEP).\index{equivalence principle!Einstein}

Einstein noted that if all bodies fall in the same way in
an external gravitational field, an observer in freefall
will find that freely falling bodies in his or her
neighborhood move with uniform velocities relative to him
or her and that the physics of pure particle mechanics in
that neighborhood is indistinguishable from mechanics in the
absence of gravity.  This led him to suggest that a freely
falling observer might find all other nongravitational physics
in his or her neighborhood to be indistinguishable from such
physics in the absence of gravity.  Einstein then proceeded to
show that if this were true for electrodynamic physics, which
was all fundamental physics at the time, light propagating
out of a gravitational potential well must suffer a redshift.

The following modern formulation of the EEP expresses the possibility  
suggested by Einstein in
1907.  It states that the outcome of any local, nongravitational
test experiment is independent of the experimental apparatus'
velocity relative to the gravitational field and is independent
of where and when in the gravitational field the experiment is
performed.  This captures Einstein's suggestion because, in
principle, local nongravitational test experiments can be
performed in spacetime regions where gravity is negligible.
The two conditions that the EEP
imposes are referred to as
local Lorentz and local position invariance, \index{local position invariance} \index{local Lorentz invariance} respectively.
Note that a \emph{local} experiment is one performed within a
spacetime region so small that the experimental apparatus can
detect no tidal effects.  A \emph{test} experiment is one
performed with an apparatus having a mass so small that the
apparatus can detect no effect of the perturbation it induces
in the gravitational field.

From the perspective of 1907, the EEP
is a striking generalization from the observed equality of test
body accelerations in a gravitational field, and, as a matter of
history, the gravitational redshift was the first physical
consequence to be derived on the assumption that this
generalization is valid.  Given this history, it is not
surprising that refinements of experimental tests of the
universality of freefall acceleration and of measurements of
the gravitational redshift remain among the most widely
recognized tests of the validity of the EEP.  Efforts to refine both kinds  
of test continue
today.  Readers can refer to Lute Maleki's article \cite{Maleki}
in this volume for details on a proposed space--based variation
on gravitational redshift measurements.

Tests of the universality of free fall acceleration are
often referred to as E\"otv\"os experiments because of the
classic torsion--balance version performed by Baron von
E\"otv\"os and collaborators early in this century \cite{Eotvos}.
As noted above, the process of refining such tests continues
today.  For example, the group of Adelberger at the University of
Washington \cite{EotWash} recently reported results of a
torsion--balance experiment that include the conclusion that
the gravitational accelerations of beryllium and copper test
bodies toward the Earth are equal to better than 2.5 parts
in $10^{12}$.  The ultimate refinement of tests of the
equality of such accelerations may well be represented by the
space--based STEP experiment under development at Stanford
University.  Readers can refer to an article in this volume for details on  
STEP \cite{STEP}.

Interest in such experiments remains high because our
understanding of their significance as tests of the EEP has evolved  
significantly since 1907.
The clear distinction Einstein made at that time between
particle mechanics and other nongravitational physics,
specifically electromagnetic physics, is no longer viable.
We now understand how test bodies are composed of atoms and
that they, in turn, are composites of the mass--energy of
nucleons and electrons and of the electromagnetic, weak-- and
strong--interaction binding energies of these particles.
Consequently, the freefall acceleration of test bodies can
be influenced by many, if not all, aspects of nongravitational
physics in an external gravitational field, and experiments
which test the universality of such accelerations turn out
to be more profound tests of the EEP than one could have realized in 1907.   
We will
return to this point later in section \ref{sec:THepsilonmu}.

Set aside, for a moment, the issue of whether or not the
EEP is valid and questions
regarding the precision to which we may be able to establish
experimentally that it is valid.  Is it possible to formulate
a gravitation field theory which predicts that it is valid?
It took Einstein almost ten years from the time of his 1907
insight to establish that the answer to this question is yes.
He did so by formulating general relativity \cite{Einstein2}.
Interested readers can refer to the recent review by Norton
\cite{Norton} for a discussion of the fascinating and in many
ways still controversial history of Einstein's development of
general relativity and of others' early attempts to understand
the theory.  This story is plagued by many formulations of the
equivalence principle, confusion regarding the significance
of coordinates and covariance and so on.  We will not delve
further into it here.
Instead, we discuss the kind of analysis one must do to determine whether or  
not any given gravitation field theory predicts the validity of the EEP.

\subsection{Theoretical contexts for analyses of the EEP}

If an analysis reveals outcomes of some local nongravitational
test experiment that depend on the velocity of the
experimental apparatus relative to an external gravitational
field or on where or when in that field the experiment is
performed, it is clear that the underlying theory is
nonmetric, that is, it violates the EEP.  Significantly, such an analytical  
result also
provides the basis for actual experiments that search for the
specific preferred--frame or preferred--location \index{preferred frame effect} \index{preferred location effect} effect revealed
by the analysis.  This approach has led to the development
of many stringent new tests of the EEP that are quite different from  
familiar E\"otv\"os
experiments and gravitational redshift measurements.  It has
also clarified which structural features of gravitation
field theories are constrained by experimental evidence that
the EEP is valid to some level of
precision.

Lagrangian field theory provides a natural setting for a
general discussion of gravitation field theories.  However,
we note that much of what follows can be discussed in terms
of gravitation field equations and matter--field equations
of motion.  Indeed, a number of recent papers exploit this
latter approach to consider modifications of the Maxwell
equations caused by quantum gravity  \index{quantum gravity} effects.  Papers by
Gambini and Pullin \cite{GP} and by Ellis \emph{et al}.
\cite{Ellis} are examples.  The paper of Haugan and
L\"ammerzahl \cite{HL} begins the analysis of physical
consequences of a broad range of conceivable Maxwell equation
modifications of this kind and the consideration of
experiments that could detect them or constrain their magnitudes.

Returning to Lagrangian--based gravitation theories, we recall
that each admits a formulation via an action principle,
\begin{equation}
\delta \! \int \! {\cal L} (\psi_g , \psi_m ) \; d^4x = 0.
\end{equation}
Here, $\psi_g$ denotes dependence of the Lagrangian density
on gravitational potentials and their derivatives, and
$\psi_m$ denotes dependence on matter fields and their
derivatives.  Not that long ago, one would have restricted
attention to theories in which field derivatives appear
only in conventional ``kinetic'' terms in ${\cal L}$, but
attitudes have changed so that field theories tend to be
viewed as effective rather than fundamental theories.
Consequently, there is now a greater willingness to consider
dependence on higher--order derivatives and the presence of
derivative couplings between fields.  Such things can make
theories nonrenormalizable, but this is not the issue for
effective theories that it is for fundamental ones.

A theory's Lagrangian density ${\cal L}$ can be split into
a purely gravitational part and ``nongravitational'' remainder,
${\cal L} = {\cal L}_g + {\cal L}_{ng}$.  The gravitational
part ${\cal L}_g$ depends only on gravitational potentials and
their derivatives.  Its form specifies the dynamics of free
gravitational fields in the theory.  The nongravitational
part ${\cal L}_{ng}$ depends on gravitational potentials and
their derivatives and on matter fields and their
derivatives.  Its form specifies the coupling between matter
and gravity in the theory.  Its form determines both how
matter responds to gravity and how matter acts as a source
of gravity.

\subsection{The role of locality}

The matter fields involved in a local, nongravitational test
experiment do not perturb gravitational potentials to a
degree that the experiment can detect.  It follows that when
using a theory to predict the outcome of such an experiment
we can treat gravitational potentials as specified functions
which represent the gravitational environment generated by some
source.  They are external potentials.  To predict
the outcome of the experiment one needs only a theory's
equations which govern the evolution of matter fields in the
relevant gravitational environment.  We derive these
gravitationally--modified equations of motion from the
theory's action principle (1) by considering variations of
the matter fields while keeping the external gravitational
potential functions fixed in the appropriate form.
Consequently, the outcomes of experiments that directly
test the EEP depend on the
form of a theory's nongravitational Lagrangian density
${\cal L}_{ng}$ alone.  Clearly, evidence that the
EEP is valid to some precision
can constrain only the manner in which matter couples to
gravity.

Since experiments that directly test the EEP are local as well as test  
experiments,
their outcomes are insensitive to the global form of
external gravitational potentials.  It is sufficient to
consider initial terms of the Taylor--series expansions of
external potentials when using a theory's action principle
to predict a local experiment's outcome.  The expansions
should be made about an event inside the experimental
apparatus during the course of the experiment.  Keeping
terms through first order is sufficient to predict the
outcome of any local experiment.  However, in some
important cases it is sufficient to keep only zeroth--order
terms.  In particular, this can be the case when an experiment
is completed quickly enough that the experimental apparatus
can detect no effect of external potential time dependence
and no effect of accelerations induced by external potential
spatial dependence.  This is generally true of realistic
local experiments which measure atomic transition frequencies,
for example, and explains why atomic clocks may generally be 
treated as realizations of ideal clocks in the sense defined by 
theories of relativity.  

To make the remainder of this discussion a bit more concrete
focus on experiments involving such atomic transitions.  To 
this point, we have established that a 
theory's predictions of their outcomes follows from
the form that its nongravitational Lagrangian density
${\cal L}_{ng}$ takes when values of the external gravitational
potentials and, if there are derivative matter--gravity couplings,
values of their derivatives at an event inside the experimental
apparatus during the experiment are plugged in.  The resulting
Lagrangian density has no explicit dependence on the spacetime
coordinates and involves only matter fields.  It determines the
gravitationally--modified equations of motion which govern
the structure of atoms treated as local test bodies.  Note,
however, that if we are interested only in atomic transition
frequencies, we need not deal with these equations.  We can,
instead, compute energies of atomic states directly.

\subsection{Relevant observables}

An expression for such energies follows from the form of
the Lagrangian density introduced in the preceding paragraph
because it is time independent.  If \emph{natural}
coordinates are used in representing a theory's action
principle (1), the energy expression's form will be that
of the Standard Model Hamiltonian plus perturbating terms.
In this context, natural coordinates are ones in which the
form of the representation of the theory's nongravitational
Lagrangian density ${\cal L}_{ng}$ reduces to the familiar
representation of the Standard Model Lagrangian density as
gravity is ``turned off.''  Schwarzschild coordinates
provide a familiar example of natural coordinates in the
context of general relativity and situations in the
external gravitational potential is static and
spherically symmetric.

Accurate estimates of the energies of atomic states are
easily computed when the gravitationally--modified
Hamiltonian is a perturbed Standard Model Hamiltonian.  In
general, the results depend on an atom's velocity through and
location in the external gravitational potentials.  This is the
case because the perturbing Hamiltonian terms reflect not only
the form of the gravitation theory's nongravitational Lagrangian
density ${\cal L}_{ng}$ but also the atom's gravitational
environment.  This environment is represented by the values
in the atom's neighborhood of the external gravitational potentials
and, if there are derivative matter--gravity couplings, the values
of their derivatives.

Despite their velocity and location dependence these computed
atomic state energies may not represent preferred--frame or
preferred--location effects that signal violation of the Einstein
equivalence principle.  Even when we are using natural coordinates
they are merely \emph{coordinate} energies.  Only velocity or
location dependence of an experimentally measured atomic state
energy, or frequency of a transition between such states,
would constitute a genuine preferred--frame or
preferred--location effect.

The distinction between coordinate energies or frequencies
and measured energies or frequencies is, in some respects,
subtle.  However, it is not difficult to appreciate if
one remembers that, fundamentally, any
measurement is simply the comparison of a property of some
system of interest to the corresponding property of a chosen
standard system.  Thus, a measurement of the frequency of
a transition between some pair of atomic states is simply a
comparison of its coordinate frequency to the coordinate
frequency of a selected standard transition.  For example,
one can imagine locking a laser to the transition whose
frequency is to be measured and a second laser to the frequency
of the standard transition.  The coordinate frequency of each
laser can depend on velocity through or location in
the external gravitational potential, but the relative or
beat frequency between them, which is the measured frequency,
may not.  While it may seem far--fetched to imagine cases
in which coordinate energies of all atomic states depend
on velocity through or location in external gravitational
potentials in precisely the same way, thus, causing such
dependence to cancel from measured energies, this is precisely
what metric theories of gravity like general relativity predict.  
The preferred--frame or preferred--location effects
predicted by nonmetric theories of gravity occur because their
matter--gravity couplings distinguish between the contributions
of rest--mass and of different types of interaction--energy to
the energies of atomic states and, so, prevent such universal 
cancellations of coordinate effects.  If such couplings are 
present, the observable frequency ratio for a pair of atomic 
clocks whose ticking rates are governed by different atomic 
transition and which move together through an external gravitational 
potential can depend on the clocks' location in and velocity 
through the potential.  In the limit of slow motion and weak gravitation 
this frequency ratio takes the form 
\begin{equation}
\frac{\nu_1({\vec x}, {\vec v})}{\nu_2({\vec x}, {\vec v})} =  
\frac{\nu_1^0}{\nu_2^0} \left(1 + \alpha_{ij} v^i v^j + \beta_{ij}  
U^{ij}({\vec x}) \right) , \label{FrequencyRelation}
\end{equation}
where the $x^i$ denote natural spatial coordinates that reduce to 
Cartesian ones as gravity is turned off, the $v^i$ denote corresponding 
components of the clocks' coordinate velocity and $U^{ij}({\vec x})$ 
denotes the usual Newtonian gravitational potential tensor at the 
clocks' location.  The parameters $\alpha_{ij}$ and $\beta_{ij}$ 
depend on the particular transitions controlling the atomic clock 
rates, except in metric theories of gravity which predict that they 
all vanish.  Their tensor character reflects the fact that the 
orientation of the atoms whose transitions govern the atomic clock 
rates can affect the observed frequency ratio (2).  Nonvanishing 
$\alpha_{ij}$ and $\beta_{ij}$ parameters characterize 
preferred--frame and preferred--location effects, respectively.   

To conclude this discussion of the way in which Lagrangian--based
gravitation theories predict outcomes of local, nongravitational
test experiments, briefly consider the acceleration of test
bodies in an external gravitational field once more.  We focus
on the freefall of atoms since, as we noted earlier, realistic
test bodies are simply assemblages of them.

The analysis of atomic systems outlined above yields an
expression for the coordinate energy of any atom in any
state of interest.  This energy is a function of the atom's
velocity relative to and location in an external gravitational
potential
\begin{equation}
E = m c^2 + \frac{1}{2} m \left(\delta_{ij} + \frac{\delta  
m_{\hbox{\scriptsize i}ij}}{m}\right) v^i v^j + m \left(\delta_{ij} +  
\frac{\delta m_{\hbox{\scriptsize g}ij}}{m}\right) U^{ij}(\vec{x}) \, ,  
\label{HLenergy}
\end{equation}
where $\delta m_{\hbox{\scriptsize i}ij}$ and $\delta m_{\hbox{\scriptsize  
g}ij}$ are the anomalous inertial and gravitational mass tensors. \index{inertial mass}\index{gravitational mass} They 
depend on  the particular state of the particular atom under 
consideration, except in metric theories of gravity which predict that 
they vanish.  The $\alpha_{ij}$ and $\beta_{ij}$ parameters 
appearing in  (\ref{FrequencyRelation}) are determined by the 
anomalous mass tensors of the states involved in the atomic clock 
transitions considered above. 

When the external gravitational potential is time
independent, the coordinate energy function (\ref{HLenergy}) is
globally conserved and its dependence on atomic velocity and 
location determines the atom's coordinate acceleration via familiar 
energy conservation arguments, 
\begin{equation}
a^i = \delta^{ij} \partial_j U + \frac{\delta m_{\hbox{\scriptsize  
i}}^{ij}}{m} \partial_j U + \delta^{ij} \frac{\delta m_{\hbox{\scriptsize  
g}kl}}{m} \partial_j U^{kl}(\vec{x}),  \label{acceleration}
\end{equation}
(here $\delta  m_{\hbox{\scriptsize i}}^{ij} = 
\delta m_{\hbox{\scriptsize i}ij}$).  Kenneth  Nordt\-vedt \cite{Nordt}
and Mark Haugan \cite{H79} exploit such arguments to relate the
outcomes E\"otv\"os experiments to the outcomes of gravitational
redshift measurements and other tests of the EEP.

In the end, the preceding overview of the kind of analysis one
must do to determine the outcomes of local nongravitational
test experiments predicted by Lagrangian--based gravitation field
theories brings one full circle.  We have come back
to the most familiar experimental tests of the EEP, but with a deeper  
appreciation of the significance of their results.  In the next 
section we consider examples of nonmetric theories and formalisms 
encompassing whole classes of such theories within which 
preferred--frame and preferred--location effects have been analyzed to 
provide a basis for testing the EEP.  Much of the work on tests of 
the EEP done before 1993 is thoroughly reviewed in  the early chapters 
of Clifford Will's \emph{Theory and Experiment in  Gravitation Physics} 
\cite{TEGP}, for an update see \cite{Will98}, see also  
\cite{HauganWill87}.

\section{Theoretical frameworks for the analysis of EEP tests}

The approach outlined in the preceding section can be used to 
determine prefer\-red--frame and preferred--location effects predicted 
by any nonmetric theory of gravity.  Such effects reflect the form 
of the theory's nongravitational Lagrangian density ${\cal L}_{ng}$ 
once the external gravitational potential in which a local 
nongravitational test experiment is performed has been plugged in.  
Proceeding in this fashion, we would have to analyze and reanalyze 
any given experiment to determine its outcome as predicted by competing 
theories.  

It is, instead, more efficient to analyze local nongravitational test 
experiments once and for all within the context of more general 
theoretical frameworks that encompass broad classes of nonmetric 
gravitation theories and gravitational environments.  Such frameworks 
are based on models of the nongravitational Lagrangian or corresponding 
matter field equations that depend on phenomenological gravitational 
potentials in ways that encompass the forms of these structures  
in many nonmetric theories and environments.  The outcome 
of an experiment predicted 
within such a framework immediately yields the outcome predicted by 
any nonmetric theory it encompasses when the framework's 
phenomenological gravitational potentials are expressed in terms of 
the particular theory's potentials.  The efficiency of this 
approach is somewhat like that provided by the PPN formalism 
\cite{TEGP} when dealing with the gravitational dynamics of metric 
theories.  

Analyses carried out in the general nonmetric frameworks discussed 
below have the additional benefit of identifying mechanisms that 
lead to preferred--frame or preferred--location effects in entire 
classes of nonmetric theories and of providing theory--independent 
parametrizations of such effects that are useful in discussing 
the results of experiments designed to search for them.  

\subsection{The $TH\epsilon\mu$--formalism}\label{sec:THepsilonmu}\index{THepsilonmu@$TH\epsilon\mu$--formalism}

The $TH\epsilon\mu$--formalism (see \cite{LightmanLee73,TEGP}) is  
based on the form of the Lagrangian governing the dynamics of point 
particles with mass $m_i$ and charge $q_i$ and of the electromagnetic 
field in a static, spherically symmetric background gravitational field 
described by the phenomenological gravitational potentials 
$T$, $H$, $\epsilon$ and $\mu$:  
\begin{equation}
L = - \sum_i m_i \int \sqrt{T - H \dot x_i^2}\, dt + \sum_i q_i \int A_a \dot  
x_i\, dx^a + \frac{1}{8\pi} \int \left(\epsilon {\mbox{\boldmath$E$}}^2 -  
\frac{1}{\mu} {\mbox{\boldmath$B$}}^2\right) d^4 x.  
\end{equation}
A striking feature of this framework and the nonmetric theories it 
encompasses is that the limiting speed of massive particles in the 
neighborhood of some point in the gravitational field can differ from 
the speed of light there.  These coordinate speeds are given, 
respectively, by the values of $\sqrt{T/H}$ and $1/\sqrt{\epsilon\mu}$ 
at the point of interest.  Preferred--frame effects result when the ratio 
of these speeds is not unity.  Variation of the relative values of 
$T$, $H$, $\epsilon$ and $\mu$ with position in the gravitational field 
can also lead to preferred--location effects.  Computations of the 
energies of atomic states using natural quantum mechanical extensions 
of the classical $TH\epsilon \mu$ Lagrangian reveal both kinds of 
effects and yield predictions for anomalous inertial and gravitational 
mass tensors (compare Eq.(\ref{HLenergy})) \cite{H79,Will74,GabrielHaugan90}.  
A quantum field theoretic extension of the formalism reveals EEP 
violations discernable in measurements of the Lamb shift, the anomalous 
magnetic moment of the electron and related phenomena 
\cite{AlvarezMann96a}.

This test theory has been widely used to interpret the results of 
experimental tests of the EEP.  For example, its predictions of 
the energies of atomic states \cite{H79}, \cite{Will74} and 
\cite{GabrielHaugan90} have be used to interpret Hughes--Drever type 
experiments as well as the Vessot--Levine \index{Vessot--Levine experiment} rocket redshift experiment 
\cite{Turneaureetal83,TEGP}.  

Originally conceived as a framework for analyzing the physics of 
charged particles and electromagnetic fields in an external 
gravitational field, the $TH\epsilon\mu$--formalism has also been 
extended in a natural way to cover the other sectors of 
nongravitational physics comprising the Standard Model \index{Standard Model} 
\cite{Horvathetal88}.  

\subsection{The $\chi g$--formalism}

Like the $TH\epsilon\mu$--formalism the $\chi g$--formalism introduced 
by W.--T.\ Ni \cite{Ni77} originally provided a framework for the 
analysis of electrodynamic physics in a background gravitational 
field and has subsequently been extended to cover other sectors of 
the Standard Model.  Unlike the $TH\epsilon\mu$--formalism the 
$\chi g$--formalism is not restricted to static, spherically symmetric 
gravitational environments.  The $\chi$ of its name refers to a 
tensor field appearing in the electromagnetic part of the 
nongravitational Lagrangian density upon which the formalism is based, 
\begin{equation}
{\cal L}_{\rm em} = - {1 \over{16\pi}} \chi^{\alpha \beta \gamma \delta}
F_{\alpha \beta} F_{\gamma \delta}. \label{chi} 
\end{equation}
The independent components of this tensor comprise twenty--one 
phenomenological gravitational potentials capable of representing 
gravitational fields in a very broad class of nonmetric 
gravitation theories.  

The coupling of one particular phenomenological potential to the 
electromagnetic field is interesting because it can be expressed 
as a purely derivative coupling to a pseudoscalar field $\varphi$.  
A particle physicist would describe it as an axion coupling.  A relativist 
would describe it as a coupling to axial torsion \cite{PLH97}.  The 
Hojmann--Rosenbaum--Ryan--Shepley theory  \cite{Hoimannetal78} is but 
one example of a theory encompassed by Ni's $\chi g$--formalism.  It 
features a novel torsion coupling that has been shown to predict 
effects inconsistent with the results of experimental tests of the 
weak equivalence principle.  

\subsection{The Kostelecky Formalism} 

String theory has the potential to provide a quantum theory of gravity 
that is unified with other fundamental theories of matter and 
interactions.  Recently Colladay and Kostelecky have introduced 
a framework for treating the possibility of spontaneous breakdown 
of Lorentz symmetry in the context of string theory  
\cite{ColladayKostelecky97,ColladayKostelecky98}.  While somewhat 
different from the sources of preferred--frame effects considered to 
this point, these string induced effects are considered here because 
they lead to modifications of the Dirac and Maxwell equations like 
those considered in the next subsection.  

\subsection{Formalisms based on matter--field equations of motion}

The effects an external gravitational field on the dynamics of matter 
fields can be dealt with at the level of equations of motion rather 
than Lagrangians.  A formalism based on forms of the equations of 
motion has the advantage of directly addressing the following 
natural requirements one would demand of the dynamics of any 
quantum field (i) deterministic evolution, (ii) the superposition 
principle, (iii) a finite propagation speed (whose maximum value, 
since it need not be isotropic, we call $c_D$) and  (iv) the 
conservation of probability.  

The equations governing the motion of a Dirac field which satisfy 
these requirements are a first--order hyperbolic system of the 
form 
\begin{equation}
0 = i \widetilde\gamma^\mu \partial_\mu \varphi + M \varphi\, ,
\end{equation}
or, in $3 + 1$--form ($\hat\mu = 1, 2, 3$)
\begin{equation}
i \partial_0 \varphi = c_D \widetilde\alpha^{\hat\mu} c \partial_{\hat\mu}  
\varphi + c_D \widetilde\Gamma \varphi + m c_D^2 \widetilde\beta \varphi \label{GDE3+1}
\end{equation}
which we call a generalized Dirac equation ($\widetilde\alpha^{\hat\mu} =  
(\widetilde\gamma^0)^{-1} \widetilde\gamma^{\hat\mu}$).
The matrices $\widetilde\gamma^\mu$ are not assumed to define a  
Clifford algebra, instead they satisfy 
$\widetilde\gamma^\mu \widetilde\gamma^\nu +  
\widetilde\gamma^\nu \widetilde\gamma^\mu = 2 g^{\mu\nu} + X^{\mu\nu}$ 
where  $g^{\mu\nu} = \frac{1}{4} \hbox{tr}(\widetilde\gamma^\mu  
\widetilde\gamma^\nu)$ and $X^{\mu\nu}$ is a matrix.
In general, $M$ is also a matrix.  A distinctive feature of this 
generalized Dirac equation is that it predicts a splitting of the 
null cones and mass shells. (For another modification of the Dirac equation see \cite{LaemmerzahlBorde00} in this volume.)

Taking the non--relativistic limit and specifying a general  
position--depen\-dence of the matrices $\widetilde\gamma^\mu$ and $M$, one  
derives the generalized Pauli equation \cite{Laemmerzahl98}
\begin{eqnarray}
i \frac{\partial}{\partial t} \varphi & = & - \frac{1}{2m} \left(\delta^{ij}  
- \frac{\delta m_{\hbox{\scriptsize i}}^{ij}}{m} - \frac{\delta \bar  
m_{\hbox{\scriptsize i} k}^{ij} \sigma^k}{m}\right) \partial_i \partial_j  
\varphi + \left(c_D A^i_j + \frac{1}{m} a^i_j\right) \sigma^j i \partial_i  
\varphi \label{GPE}\\
& & + \Bigl[m\, U(\mbox{\boldmath$x$}) +  
\mbox{\boldmath$C$}\cdot\mbox{\boldmath$\sigma$}\, m\, U(\mbox{\boldmath$x$})  
+
\delta m_{\hbox{\scriptsize g}ij}
U^{ij}(\mbox{\boldmath$x$}) + c_D \,  
\mbox{\boldmath$T$}\cdot\mbox{\boldmath$\sigma$} + m c_D^2  
\mbox{\boldmath$B$}\cdot\mbox{\boldmath$\sigma$}\Bigr] \varphi \nonumber
\end{eqnarray}
where the anomalous coefficients $\delta m_{\hbox{\scriptsize i}}^{ij}$,  
$\delta \bar m_{\hbox{\scriptsize i} k}^{ij}$, $A^i_j$, $a^i_j$,  
$\mbox{\boldmath$C$}$, $\delta m_{\hbox{\scriptsize i}}^{ij}$,  
$\mbox{\boldmath$B$}$ stem from those parts of the  
$\widetilde\gamma$--matrices which prevent them from defining a 
Clifford algebra and  from anomalous terms in the mass matrix $M$, 
for example, $\frac{\delta  
m_{\hbox{\scriptsize i}}^{ij}}{m} + \frac{\delta \bar m_{\hbox{\scriptsize i}  
k}^{ij}}{m} \sigma^k = \frac{1}{2} (1 + \beta) \widetilde\alpha^{(i}  
\widetilde\alpha^{j)}$ where $\beta$ is the usual Dirac $\gamma^0$ and  
$\sigma^i$ are the usual Pauli matrices.  The generalized Pauli 
equation predicts preferred--frame and preferred--location effects.  
Terms like those representing couplings between spin and the Newtonian 
gravitational potential were first introduced in references 
\cite{HDass76,HDass77} and \cite{Peres78}.

The Pauli equation (\ref{GPE}) is a generalization of Schroedinger 
equation provided by M . Haugan's approach \cite{H79} to the dynamics 
of scalar matter.  As in the case of preceding formalisms, this Pauli 
equation provides a basis for broad range of experimental tests of the 
EEP, including experiments exploiting matter--wave interferometry.  The 
classical limit of the generalized Pauli equation describes the free 
fall of classical spin--polarized bodies,
\begin{equation}
a^i = \delta^{ij} \partial_j U + \left[\frac{\delta m_{\hbox{\scriptsize  
i}}^{ij}}{m} +  2 \left({{\delta\bar m_{\hbox{\scriptsize i} k}^{ij}}\over m}  
+ \delta^{ij} C_k\right) S^k\right] \partial_j U + \delta^{ij} \, \frac{\delta  
m_{\hbox{\scriptsize g}kl}}{m}\,  \partial_j U^{kl}(\vec{x}) \, .  
\label{extaccel}
\end{equation}
Notice that not all of the anomalous parameters appearing in the quantum 
equation (\ref{GPE})  survive in the classical freefall acceleration.  
Only by considering the evolution of the spin as well can one design 
experiments in the classical limit that are sensitive to all possible 
anomalies.  

Once a generalized Dirac equation (\ref{GDE3+1}) is available we can address 
the dynamics of the electromagnetic field in an analogous way.  The 
electromagnetic field can be defined operationally by considering the  
phase shifts in charged particle interferometry.  Assuming that the 
dynamics of electromagnetic fields satisfies the same requirements as we 
demanded for the dynamics of the Dirac field, this leads to generalized Maxwell 
equations of the form,  
\begin{equation}
\partial_{[\mu} F_{\nu\rho]} = 0 \, , \qquad 4 \pi j^\mu = \lambda^{\mu\nu\rho\sigma}  
\partial_\nu F_{\rho\sigma} + \bar\lambda^{\mu\rho\sigma} F_{\rho\sigma}  
\, . \label{GME}
\end{equation}
In the case of small deviations from minimal coupling to the 
Riemannian space--time metric $g_{\mu\nu}$, we have  
$\lambda^{\mu\nu\rho\sigma} = \delta^{\mu[\rho} g^{\sigma]\nu} +  
\delta\lambda^{\mu\nu\rho\sigma}$ with small values of 
$\delta\lambda^{\mu\nu\rho\sigma}$ and $\bar\lambda^{\mu\rho\sigma}$. 
Clearly, $\delta\lambda^{\mu\nu\rho\sigma}$ can induce 
anistropic propagation of light and birefringence. The 
$\bar\lambda^{\mu\rho\sigma}$ can also modify propagation, in some 
cases leading to a damping of electromagnetic waves.  

The generalized Dirac equation (\ref{GDE3+1}) and Maxwell equations (\ref{GME}) 
can be used just as the corresponding equations that emerge from the 
$TH\epsilon\mu$--formalism or the $\chi g$--formalism, respectively, to analyze to 
properties of atoms in background gravitational fields.  They do, 
however, encompass a wider range of nonmetric couplings that 
influence spin and polarization.  Consequently, they provide the 
broadest possible basis for the interpretation of experimental tests 
of the EEP.  

\section{Motivations for continued testing of the EEP}

Although all tests of the EEP, including some of remarkable precision, 
have so far failed to detect any hint of a violation, recent theoretical 
developments continue to suggest the EEP must be violated at some level.  
All approaches to quantizing gravity and to unifying it with the other 
fundamental interactions currently under study are capable of predicting 
such violations.  

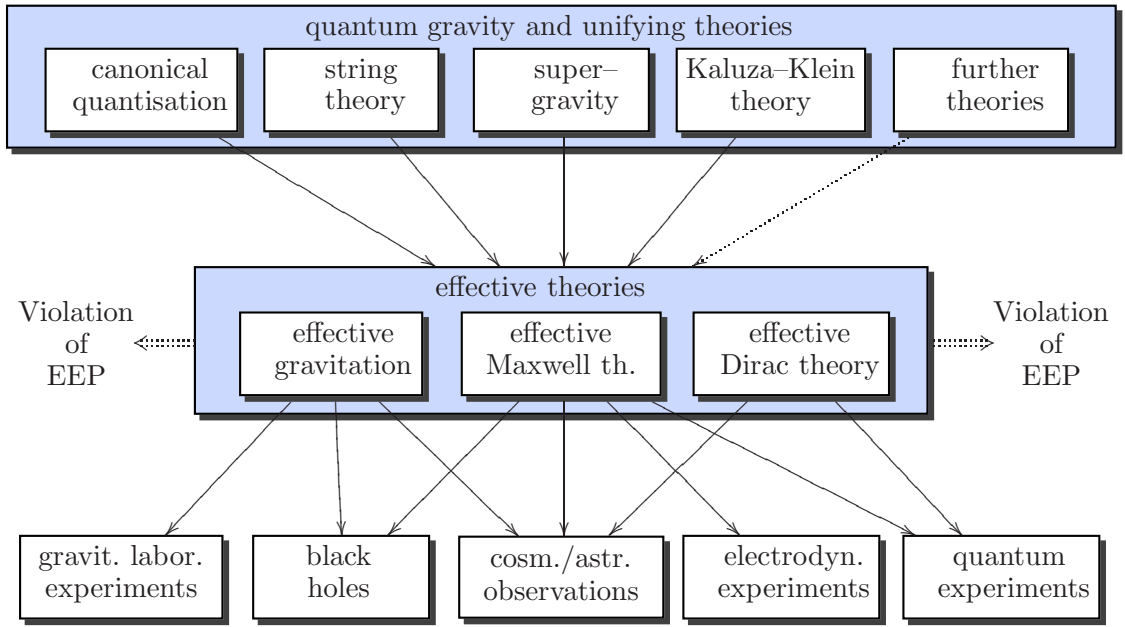
\begin{figure}[t]
\begin{equation*}
\begin{xy} 0;<0.9cm,0cm>:
,(0,7.8)*{\ShowX{light-blue}{14.5cm}{\begin{center}\txt{\qquad\qquad\qquad\qquad\qquad quantum gravity and  
unifying theories}\end{center}\vspace*{1cm}}}
,(-6.2,7.6)*{\ShowX{white}{1.9cm}{\begin{center}\txt{\; canonical \\  
\; quantisation}\end{center}}}="KQ"
,(-3.1,7.6)*{\ShowX{white}{1.9cm}{\begin{center}\txt{\quad\textcolor{white}{K}string \textcolor{white}{K}  
\\ \quad \textcolor{white}{g}theory\textcolor{white}{g}}\end{center}}}="ST"
,(0,7.6)*{\ShowX{white}{1.9cm}{\begin{center}\txt{\quad\textcolor{white}{K}super--\textcolor{white}{K}  
\\ \quad\textcolor{white}{g}gravity\textcolor{white}{g}}\end{center}}}="SG"
,(3.1,7.6)*{\ShowX{white}{1.9cm}{\begin{center}\txt{Kaluza--Klein \\  
\textcolor{white}{g}theory\textcolor{white}{g}}\end{center}}}="KK"
,(6.2,7.6)*{\ShowX{white}{1.9cm}{\begin{center}\txt{\quad further \\  
\quad \textcolor{white}{g}theories\textcolor{white}{g} }\end{center}}}="WT"
,(0,3.9)*{\ShowX{light-blue}{9.5cm}{\begin{center}\txt{\qquad\qquad\qquad\qquad effective  
theories}\end{center}\vspace*{1.1cm}}}="ET"
,(-3.4,3.7)*{\ShowX{white}{2.1cm}{\begin{center}\txt{\;\; effective \\  
\;\; gravitation}\end{center}}}="EG"
,(0,3.7)*{\ShowX{white}{2.1cm}{\begin{center}\txt{effective \\  
\textcolor{white}{y}Maxwell th.\textcolor{white}{y}}\end{center}}}="EM"
,(3.4,3.7)*{\ShowX{white}{2.1cm}{\begin{center}\txt{\; effective \\  
\; Dirac theory}\end{center}}}="ED"
,(-7.2,3.9)*{\txt{Violation \\ of \\ EEP}}="EEP1"
,(7.2,3.9)*{\txt{Violation \\ of \\ EEP}}="EEP2"
,(-6.5,0.4)*{\ShowX{white}{2.1cm}{\begin{center}\txt{gravit.~labor. \\  
\textcolor{white}{p}experiments\textcolor{white}{p}}\end{center}}}="LE"
,(-3.25,0.4)*{\ShowX{white}{2.1cm}{\begin{center}\txt{\quad black \\  
\quad \textcolor{white}{p}holes\textcolor{white}{p}}\end{center}}}="GP"
,(0,0.4)*{\ShowX{white}{2.1cm}{\begin{center}\txt{cosm./astr. \\  
\textcolor{white}{y}observations\textcolor{white}{y}}\end{center}}}="KB"
,(3.25,0.4)*{\ShowX{white}{2.1cm}{\begin{center}\txt{\;\; electrodyn. \\  
\;\; experiments}\end{center}}}="SO"
,(6.5,0.4)*{\ShowX{white}{2.1cm}{\begin{center}\txt{\;\;\textcolor{white}{l}quantum\textcolor{white}{l}  
\\ \;\; experiments}\end{center}}}="QM"
%
%
%
%
\ar@2{.>} "ET";"EEP1"
\ar@2{.>} "ET";"EEP2"
\ar@{->} "KQ";"ET"
\ar@{->} "ST";"ET"
\ar@{->} "SG";"ET"
\ar@{->} "KK";"ET"
\ar@{.>} "WT";"ET"
\ar@{->} "EG";"KB"
\ar@{->} "EG";"GP"
\ar@{->} "EM";"KB"
\ar@{->} "EM";"SO"
\ar@{->} "EM";"GP"
\ar@{->} "EG";"LE"
\ar@{->} "EM";"QM"
\ar@{->} "ED";"KB"
\ar@{->} "ED";"QM"
%
\end{xy}
\end{equation*}
\vspace*{-5mm}
\caption{Sources of violations of the EEP and classes of experiments or  
observations that are sensitive to them.}
\end{figure}

\subsection{String theory}\index{quantum gravity}\index{string theory}

Today, string theory is among the most promising candidate for a theory 
of quantum gravity fully unified with other fundamental interactions, 
and it has been shown to predict a variety of EEP violations.  

For example, departures from universal free fall accelerations have 
been computed in references \cite{DamourPolyakov94,DamourPolyakov96}. 
The composition--dependent component of test--body acceleration is 
estimated to be as large as a part in $10^{15}$ of the mean 
gravitational acceleration.  The proposed STEP experiment could 
easily detect such an anomaly (see \cite{STEP}).

String theory can also predict a time--varying fine structure  
constant because of couplings to scalar (dilatonic) fields, 
for example, see \cite{Damour99}.


In the latest versions of string theory, physical particles and fields 
are confined to the neighborhoods of $D$--branes and their propagation 
may be affected by recoil of the branes caused by that propagation 
\cite{EllisMavromatosNanopoulos99}.  The effect of this recoil 
can be accounted for via an energy--dependent effective metric.  This 
leads to modifications of the Maxwell equations, 
\begin{alignat}{2}
\mbox{\boldmath$\nabla$} \cdot \mbox{\boldmath$E$} +  
\bar{\mbox{\boldmath$u$}} \cdot \partial_t \mbox{\boldmath$E$} & = 0 & \qquad  
 \mbox{\boldmath$\nabla$} \cdot \mbox{\boldmath$B$} & = 0 \\
\mbox{\boldmath$\nabla$} \times \mbox{\boldmath$B$} - (1 - \bar u^2)  
\partial_t \mbox{\boldmath$E$} + \bar{\mbox{\boldmath$u$}} \times \partial_t  
\mbox{\boldmath$B$} + (\bar{\mbox{\boldmath$u$}} \cdot  
\mbox{\boldmath$\nabla$}) \mbox{\boldmath$E$} & = 0 & \qquad  
\mbox{\boldmath$\nabla$} \times \mbox{\boldmath$E$} & = - \partial_t  
\mbox{\boldmath$B$} \,,
\end{alignat}
which predict dispersive light propagation.  Analogous modifications 
of the Dirac equation account for the effect of brane recoil on the 
propagation of neutrinos and other fermions \cite{Ellisetal99}, 
\begin{equation}
\gamma^a \partial_a \psi - m \psi + \gamma^0 (\bar{\vec u} \cdot  
\vec{\nabla}) \psi = 0 \, .
\end{equation}

\subsection{Loop quantum gravity}\index{quantum gravity}\index{loop gravity}

In the nonperturbative approach to quantum gravity based on observables 
analogous to Wilson loops, the semi--classical gravitational field is 
described via expectation values in so--called ``weave''states.  Gambini 
and Pullin \cite{GP} discuss the propagation of light through a 
gravitational field represented by a parity--violating weave state and 
find a polarization dependence of light propagation inconsistent with 
the EEP.  The weave state is characterized by the length scale $L$, and 
gives rise to effective Maxwell equations \cite{GP} of the form 
\begin{align}
\partial_t \mbox{\boldmath$E$} & = - \mbox{\boldmath$\nabla$} \times  
\mbox{\boldmath$B$} + 2 \chi \ell_{\rm P} {\mbox{\boldmath$\nabla$}}^2  
\mbox{\boldmath$B$} \\
\partial_t \mbox{\boldmath$B$} & = \mbox{\boldmath$\nabla$} \times  
\mbox{\boldmath$E$} - 2 \chi \ell_{\rm P} {\mbox{\boldmath$\nabla$}}^2  
\mbox{\boldmath$E$} \, . \label{GP2}
\end{align}
A corresponding effective Dirac equation \cite{AlfraoMoralesTecotlUrrutia00} 
has the form 
\begin{equation}
\left(i \widetilde\gamma^a \partial_a - \widetilde m + \widetilde\gamma^{ab}  
\partial_a \partial_b\right) \psi = 0
\end{equation}
with $\widetilde\gamma^a = \gamma^a + \kappa_1 \frac{\ell_{\rm P}}{L} G_1^a +  
\kappa_2 \left(\frac{\ell_{\rm P}}{L}\right)^2 G_2^a + \ldots$, $\widetilde m = m +  
\lambda_1 \frac{\ell_{\rm P}}{L} M_1 + \lambda_2 \left(\frac{\ell_{\rm P}}{L}\right)^2 M_2 +  
\ldots$, and $\widetilde\gamma^{ab} = \mu_1 \frac{\ell_{\rm P}}{L} G_1^{ab} + \mu_2  
\left(\frac{\ell_{\rm P}}{L}\right)^2 G_2^{ab} + \ldots$, where $\gamma^a$ are the  
usual Dirac $\gamma$--matrices, $m$ is the usual mass of the Dirac particle,  
$G_i^a$, $M_i$, and $G_i^{ab}$ are arbitrary matrices, $\ell_{\rm P}$ is the Planck  
length and the coefficients $\kappa_i$, $\lambda_i$ and $\mu_i$ are of the  
order unity ($i = 1, 2, \ldots$).

Since these field equations feature second--order spatial derivatives they 
are no longer hyperbolic and clearly single out a preferred frame.  In 
addition, note that equation (\ref{GP2}) modifies the homogeneous 
Maxwell equations, disrupting the relationship between field and 4--vector 
potential!  

\subsection{Gauge theories of gravity and other possibilities}

Gauge theories of gravity like the Poincar\'e gauge theory \index{Poincaregaugehtheory@Poincar\'e gauge theory} 
\cite{HehlHeydeKerlick76} that leads to a Riemann--Cartan geometry, or 
the gauge theory of a linear group that leads to a metric--affine theory  
\cite{Hehletal95}, gives rise to additional gravitational fields like 
torsion  and, in the latter case, to nonmetricity.  If these 
additional fields couple directly to matter, they can break local 
Lorentz invariance by singling out a preferred frame as well as 
breaking local position invariance.  

In \hfil supergravity theories, \hfil which gauge the \hfil super--Poincar\'e group, 
\hfil torsion \break emer\-ges as a bilinear combination of fundamental 
spin--$\frac{3}{2}$--field, see reference \cite{vanNieuwenhuizen82}, 
for example. \index{supergravity}

Though invented long ago, Kaluza--Klein theories \index{Kaluza--Klein theory} arise as a low--energy 
limits of string theory, with all that that entails regarding the 
validity of the EEP, see reference \cite{Duff94}, for example.  

Finally, we note that nonsymmetric theories of gravity, like those 
devised by John Moffat, have been shown to predict departures from 
universal free fall and violations of local Lorentz invariance in 
the electromagnetic sector \cite{Will89} and \cite{Gabrieletal91}.

\section{Experimental and observational tests of the EEP}

In principle, the outcomes of almost any experiment of observation 
conducted in different gravitational environments could yield 
evidence of the breakdown of  the EEP.  There are, however, 
certain classes of experiments and observations that are 
sensitive to characteristic violations of the EEP revealed by 
analyses within the theoretical frameworks discussed in section 2.  

\subsection{ Tests of the Universality of Freefall}\index{freefall}

\subsubsection{Tests with bulk matter}

Experiments that search for composition--dependence of the freefall 
acceleration of macroscopic samples of matter are direct tests of 
the weak equivalence prinicple, one consequence of the EEP.  It 
can be tested in traditional E\"otv\"os fashion using torsion 
balance technology as in reference \cite{EotWash} or by monitoring 
the relative motion of freely falling bodies as in the Bremen 
drop tower experiment \cite{Vodeletal00} and in the proposed 
MICROSCOPE \cite{Touboul00} and STEP \cite{STEP} space--based 
experiments.  To date, the highest precision, of order $10^{-12}$, 
has been achieved by torsion balance experiments, but the 
STEP experiment is designed to reach a precision of $10^{-18}$.

Equation (\ref{extaccel}) shows that macroscopic samples of 
spin--polarized \index{spin} matter may experience different gravitational 
accelerations than unpolarized matter.  Torsion balance experiments 
looking for such differential accelerations have been conducted by 
by Ritter, Gillies and coworkers 
\cite{Ritteretal90,RitterWinklerGillies93}.  They find no evidence of 
new spin--dependent forces.

\subsubsection{Tests with quantum particles}

As we saw in section 2, nonmetric theories of gravity can predict 
that quanta of different kinds fall with different accelerations 
in a gravitational field.  Historically, there has been a great deal 
of interest in direct searches for such effects, especially in 
comparing the free fall acceleration of particles and antiparticles.  

The first test of this type was performed by Witteborn and  
Fairbank \cite{WittebornFairbank67} who tried to measure the gravitational 
acceleration of charged particles.  A little later, Koester 
\cite{Koester76} showed that neutrons fall with the same way as  
classical bulk matter to an accuracy of a few percent.  This result 
has been verified by means of neutron interferometry 
\cite{Staudenmannetal80}.

The potential for future matter interferometry tests of the EEP seems 
bright.  Atomic interferometers have recently determined the 
gravitational acceleration toward the Earth to a part in $10^{9}$ and 
yield results consistent with the measured acceleration of bulk 
matter.  Refinements of these devices are expected to produce still 
more precise results and can be used to search for spin--dependent 
accelerations like those in (\ref{extaccel}).  

\subsection{Spectroscopic and atomic clock tests of the EEP}\index{clock}

As noted in preceding sections atoms are composites of the mass--energy of 
nucleons and electrons as well as of their electromagnetic and weak-- and 
strong--interaction binding energies.  Nonmetric theories whose 
matter--gravity couplings distinguish between these various contributions 
can cause not only gravitational accelerations that differ from atom to 
atom but also shifts in the energy spacings of atomic states that depend 
on an atom's velocity through or location in its gravitational environment.  
Spectroscopic and atomic clock experiments can search directly for 
these kinds of preferred--frame and preferred--location effects.  

In \index{Hughes--Drever experiment} Hughes--Drever--like experiments, for example, see Ref.\ \cite{HughesDreverExp}, one searches for relative shifts between 
the frequencies of ground--state hyperfine transitions depending 
on atomic orientation in the gravitational environment.  The 
interpretation of this type of experiment in the context of the 
$TH\epsilon\mu$--formalism is discussed in reference \cite{TEGP}.  
The interpretation in the context of the test theory of 
section 2.4 is discussed in reference \cite{Laemmerzahl98}.

Spectroscopic methods can also be used to search for the effects of 
spin--dependent and other EEP--violating effects predicted by equation 
(\ref{GPE}), for example, see \cite{Laemmerzahl97,Laemmerzahl98}.  
These techniques have been used to verify the spin--rotation 
coupling \cite{HLMashhoon88,HL_Mashhoon95} in a search for anomalous  
spin--couplings \cite{HL_Venemaetal92}.

Atomic clock technology is a particular, refined application of 
spectroscopic technique.  Experiments that monitor the relative 
rates of different types of atomic clock for dependence on the clocks' 
velocity through or location in a gravitational field provide 
another kind of spectroscopic test of the EEP.  In essence, such 
tests are either tests of relativistic Doppler shifts that are sensitive 
to the parameters $\alpha_{ij}$ in (\ref{FrequencyRelation}) or tests 
of gravitational redshifts that are sensitive to the parameters  
$\beta_{ij}$.  The difficulty of moving clocks through large 
changes in gravitational potential or at speeds approaching that of 
light limits the precision of such experiments.  However, the 
Gravity Probe A  experiment \cite{VessotLevine80} succeeded in 
imposing the constraint $|\beta_{ij}|  \leq 10^{-4}$.  A recent 
experiment employing trapped Lithium atoms moving at 6.4\% of the 
speed of light \cite{Grieseretal94} was able to impose the constraint 
$|\alpha_{ij}| \leq 10^{-6}$.  Atomic clock technology has also 
been used to constrain EEP--violating time--dependence of the 
fine--structure constant \cite{PrestageTjoelkerMaleki95}.

\subsection{EEP tests involving observations of wave propagation}\index{wave propagation}

Observations of the propagation of electromagnetic waves or other 
fields through a gravitational field are, in sense, a kind of 
experiment examing the effects of freefall.   We discuss them 
separately, however, because of the distinctive way in which 
local effects that we think of as directly signalling violations of 
the EEP are allowed to build up as waves propagate over long 
distances.  

Departures of the form of the Maxwell equations from their usual 
metric form induced by nonmetric couplings to gravity can lead to 
dispersive \index{dispersion} wave propagation or \index{birefringence} birefringence.  Analogous departures 
of the Dirac equation from its usual metric form can also lead 
to dispersive propagation and make measurements of the arrival times 
of photons and neutrinos emitted from the same astrophysical event 
a test of EEP.  The existence of very short duration events like 
supernova explosions and gamma ray bursts in combinations with 
the build up of gravitational delays over very great distances 
makes sharp tests possible.  

Recently, limits on gravity--induced dispersion of electromagnetic 
wave propagation have been inferred from observations of quasars 
and gamma ray bursters \cite{Schaefer99}.  They constrain 
\begin{equation}
\frac{c_\gamma(\omega) - c_\gamma(\omega_0)}{c_\gamma(\omega_0)}\, , \quad  
\frac{c_\gamma^+ - c_\gamma^-}{c_\gamma^+}\, ,
\end{equation}
where $c_\gamma(\omega_0)$ is the velocity of the photon for a given  
frequency $\omega$ or polarization $\pm$.  Exploiting rapid time 
variation of gamma ray bursters, Schaefer \cite{Schaefer99} is able 
to to impose sharp constraints on gravity--induced dispersion, 
$|(c_\gamma(\omega) -  c_\gamma(\omega_0))/c_\gamma(\omega_0)| 
\leq 6 \times 10^{-21}$ for $\omega  \sim 10^{18}\;\hbox{Hz}$ and 
$\omega_0 \sim 10^{19}\;\hbox{Hz}$.  See also \cite{Billetetal99} 
for implications of such data for quantum gravity models.  Data 
constraining gravity--induced differences between the speed at which 
photons and neutrinos propagate are also imposed, 
\begin{equation}
\frac{c_\gamma^\pm(\omega) - c_\nu^\pm(\omega)}{c_\gamma^\pm(\omega)} \, ,
\end{equation}
is also available.  

Finally, observations that constrain differences between the speeds 
with which light with different polarizations propagates through a 
gravitational field have been analyzed \cite{HauganKauffmann95} leading 
to the constraint 
$|(c_{\gamma}^{+} - c_{\gamma}^{-})/c_{\gamma}^{+}| \leq 10^{-28}$.


\begin{thebibliography}{8.}
\addcontentsline{toc}{section}{References}

\bibitem{Newton} I. Newton: \emph{Philosophiae Naturalis
Principia Mathematica}  (London 1686).

\bibitem{Einstein} A. Einstein: Jahrb. Radioact. Elect. \textbf{4},
411 (1907).

\bibitem{Maleki} L. Maleki: SpaceTime Mission: Clock Test of Relativity at  
four Solar Radii, this volume.

\bibitem{Eotvos} R.V. E\"otv\"os, V. Pek\'ar and E. Fekete: Beitr{\"a}ge zum  
{G}esetz der {P}roportionalit{\"a}t von {T}r{\"a}gheit und {G}ravit{\"a}t,  
{\em
Ann. Physik} \textbf{68}, 11 (1922).


\bibitem{EotWash} Y. Su, B.R. Heckel, E.G.  Adelberger, J.H. Gundlach, M.  
Harris, G.L. Smith, and H.E. Swanson: New test of the universality of free  
fall, {\it Phys. Rev.}\textbf{ D 50},
3614 (1994).

\bibitem{STEP} N. Lockerbie, J. Mester, R. Torii, S. Vitale, and P. Worden:  
STEP: A Status Report, in C. L\"ammerzahl, C.W.F. Everitt, F.W. Hehl (eds.): {\it Gyrod, Clocks, Interferometers...: Testing Relativistic Gravity in Space}, Springer--Velag 2001, p. 213.

\bibitem{Einstein2} A. Einstein: {\em Preuss. Akad. Wiss Berlin Sitzber.}
688 (1916).

\bibitem{Norton} J. Norton: General covariance and the foundations of  
general relativity: eight decades of dispute, {\em Rep. Prog. Phys.}  
\textbf{56}, 791 (1993).

\bibitem{GP} R. Gambini and J. Pullin: Nonstandard optics from quantum  
space--time, {\em Phys. Rev.} \textbf{D 59}, 124021 (1999).

\bibitem{Ellis} J. Ellis, N.E. Mavromatos and D.V. Nanopoulos: Probing  
models of quantum space--time foam, preprint gr-qc/9909085.

\bibitem{HL} M.P. Haugan and C. L\"ammerzahl: On the experimental  
foundations of the Maxwell equations, {\em Ann. Phys.} (Leipzig) {\bf 9}, 119  (2000).

\bibitem{Nordt} K. Nordtvedt: Qualitative relationship between clocks  
gravitational ``red--shift'' violatings and nonuniversality of free--fall  
rates in nonmetric theories of gravity, {\em Phys. Rev.} \textbf{D 11}, 245  
(1975).

\bibitem{H79} M.P. Haugan: Energy Conservation and the Principle of  
Equivalence, {\em Ann. Phys.} (N.Y.) \textbf{118}, 156 (1979).

\bibitem{TEGP} C.M. Will: \emph{Theory and Experiment in Gravitation
Physics}, revised edition (Cambridge University Press, Cambridge 1993).

\bibitem{Will98} C.M. Will: The Confrontation between General Relativity and  
Experiment: A 1998 Update (Lecture notes from the 1998 SLAC Summer Institute  
on Particle Physics), gr-qc/9811036.

\bibitem{HauganWill87}
M.P. Haugan and C.M. Will:
Modern tests of special relativity,
{\em Physics Today}, May 1987, p.\ 69.

\bibitem{LightmanLee73}
A.P. Lightman and D.L. Lee:
Restricted Proof that the Weak Equivalence Principle Implies the {E}instein  
Equivalence Principle, {\em Phys.\ Rev.} \textbf{D 8}, 364 (1973).

\bibitem{Will74}
C.M. Will: Gravitational red--shift measurements as tests of nonmetric  
theories of gravity, {\em Phys.\ Rev.} \textbf{D 10}, 2330 (1974).

\bibitem{GabrielHaugan90}
M.D. Gabriel and M.P. Haugan:
Testing the {E}instein {E}quivalence {P}rinciple: Atomic clocks and local  
{L}orentz invariance, {\it Phys.\ Rev.} \textbf{D 41}, 2943 (1990).

\bibitem{Turneaureetal83}
J.P. Turneaure, C.M. Will, B.F. Farrel, E.M. Mattison, and R.F.C. Vessot:  
Test of the principle of equivalence by a null gravitational red--shift  
experiment, {\em Phys.\ Rev.} \textbf{27}, 1705 (1983).

\bibitem{AlvarezMann96a}
C. Alvarez and R.B. Mann: Testing the Equivalence Principle in the Quantum  
Regime, preprint (1996), Honorable mention in the Gravity Research Foundation  
Essay Contest (and references cited therein).

\bibitem{Horvathetal88}
J.E. Horvath, E.A. Logiudice, C. Riveros, and H. Vucetich:
Einstein equivalence principle and theories of gravitation: A  
gravitationally modified standard model, {\em Phys.\ Rev.} \textbf{D 38},  
1754 (1988).

\bibitem{Ni77}
W.--T. Ni: Equivalence Principles and Electromagnetism, {\em  
Phys.\ Rev.\ Lett.} \textbf{38}, 301 (1977); {\it Bull. Am. Phys. Soc.} {\bf 19}, 655 (1974); A Nonmetric Theory of Gravity, preprint, Montana State University, Bozeman, Montana, USA (1973), http://gravity5.phys.nthu.edu.tw.

\bibitem{PLH97}
R.A. Puntigam, C. L{\"a}mmerzahl, and F.W. Hehl:
Maxwell's theory on a post--{R}iemannian spacetime and the equivalence  
principle, {\em Class.\ Qaunt.\ Grav.} \textbf{14}, 1347 (1997).

\bibitem{Hoimannetal78} S. Hojman, M.P. Rosenbaum, and L.C. Shepley: Gauge  
invariance, minimal coupling, and torsion, {\it Phys. Rev.} {\bf D 17}, 3141  
(1978).

\bibitem{LaemmerzahlBorde00}
C. L\"ammerzahl, Ch.J. Bord\'e: Testing the Dirac equation, in C. L\"ammerzahl, C.W.F. Everitt, F.W. Hehl (eds.): {\it Gyrod, Clocks, Interferometers...: Testing Relativistic Gravity in Space}, Springer--Velag 2001, p. 466.

\bibitem{Laemmerzahl98} C. L\"ammerzahl: Quantum Tests of Foundations of  
General Relativity, {\em Class. Quantum Grav.} \textbf{14}, 13 (1998).

\bibitem{HDass76}
N.D. Hari Dass: Test for {$C$}, {$P$}, and {$T$} Nonconservation in  
Gravitaton, {\em Phys.\ Rev.\ Lett.} \textbf{36}, 393 (1976).

\bibitem{HDass77}
N.D. Hari Dass: Experimental Tests for Some Quantum Effects in Gravitation,
{\em Ann.\ Physics (N.Y.)} \textbf{107}, 337 (1977).

\bibitem{Peres78}
A. Peres: Test of the equivalence principle with spin, {\em Phys.\ Rev.}  
\textbf{D 18}, 2739 (1978).

\bibitem{ColladayKostelecky97}
D. Colloday and V.A. Kostelecky: $CPT$--violation and the standard model,  
{\it Phys.\ Rev.} {\bf D 55}, 6760 (1997).

\bibitem{ColladayKostelecky98}
D. Colloday and V.A. Kostelecky: Lorentz--violating extension of the  
standard model, {\it Phys.\ Rev.} {\bf D 55}, 6760 (1997).

\bibitem{DamourPolyakov94}
T. Damour and A.M. Polyakov: The string dilaton and a least action principle,
{\em Nucl.\ Physics} \textbf{B 423}, 532 (1994).

\bibitem{DamourPolyakov96}
T. Damour and A.M. Polyakov: String Theory and Gravity, {\em Gen.\ Rel.\  
Grav.} \textbf{12}, 1171 (1996).

\bibitem{Damour99}
T. Damour: Equivalence Principle and Clocks, to appear in the {\it  
Proceedings of the 34th Rencontres de Moriond, "Gravitational Waves and  
Experimental Gravity"}, January 1999, gr-qc/9904032.

\bibitem{EllisMavromatosNanopoulos99}
J. Ellis, N.E. Mavromatos, and D.V. Nanopoulos:
Probing models of quantum space--time foam, gr-qc/9909085.

\bibitem{Ellisetal99}
J. Ellis, N.E. Mavromatos, D.V. Nanopoulos, and G. Volkov:
Gravitational--Recoil Effects on Fermion Propagation in Space-Time Foam,
gr-qc/9911055.

\bibitem{AlfraoMoralesTecotlUrrutia00}
J. Alfaro, H.A. Morales--Tecotl, and L.F. Urrutia: Quantum gravity  
corrections to neutrino propagation, {\em Phys.\ Rev.\ Lett.} \textbf{84}, to  
appear (2000).

\bibitem{HehlHeydeKerlick76}
F.W. Hehl, P. von der Heyde, G.D. Kerlick, and J.M. Nester: General relativity with spin  
and torsion: Foundations and prospects, {\em Rev.\ Mod.\ Phys.} \textbf{48},  
393 (1976).

\bibitem{Hehletal95} F.W. Hehl, J.D. McCrea, E.W.  Mielke, and Y. Ne'eman:  
Metric--affine gauge theory of gravity: Field Equations, Noether Identities,  
World Spinors, and Breaking of Dilation Invariance,
{\em Phys. Rep.} {\bf 258}, 1 (1995).

\bibitem{vanNieuwenhuizen82}
P. van Nieuwenhuizen: Supergravity, {\em Phys.\ Rep.} \textbf{68}, 189 (1982). 

\bibitem{Duff94}
M.J. Duff: Kaluza--Klein theory in perspective, hep-th/9410046.

\bibitem{Will89}
C.M. Will, C.M.: Violation of the Weak Equivalence Principle in Theories of  
Gravity with a Nonsymmetric Metric, {\it Phys.\ Rev.\ Lett.} \textbf{62}, 369  
(1989).

\bibitem{Gabrieletal91}
M.D. Gabriel, M.P. Haugan, R.B. Mann, and J.H. Palmer: Nonsymmetric  
gravitation theories and local {L}orentz invariance, {\it Phys.\ Rev.}  
\textbf{D 91}, 2465 (1991).

\bibitem{Vodeletal00}
W. Vodel, H. Dittus, S. Nietzsche, H. Koch, J. v. Zameck Glyscinski, R.  
Neubert, S. Lochmann, C. Mehls, D. Lockowandt: High Sensitive DC SQUID Based  
Position Detectors for Application
in Gravitational Experiments at the Drop  
Tower Bremen, this volume.

\bibitem{Touboul00}
P. Touboul: Space Accelerometers Present Status, in C. L\"ammerzahl, C.W.F. Everitt, F.W. Hehl (eds.): {\it Gyrod, Clocks, Interferometers...: Testing Relativistic Gravity in Space}, Springer--Velag 2001, p. 274.

\bibitem{Ritteretal90}
R.C. Ritter, C.E. Goldblum, W.-T. Ni, G.T. Gillies, and C.C. Speake:  
Experimental test of equivalence principle with polarized masses,
{\it Phys.\ Rev.} \textbf{D 42}, 977 (1990).

\bibitem{RitterWinklerGillies93}
R.C. Ritter, L.I. Winkler, and G.T. Gillies:
Search for Anomalous Spin--Dependent Forces with a Polarized--Mass Torsion  
Pendulum, {\it Phys.\ Rev.\ Lett.} \textbf{70}, 701 (1993).

\bibitem{WittebornFairbank67}
F.C. Witteborn and W.M. Fairbank:
Experimental Comparison of the Graviational Force on Freely Falling  
Electrons and Metallic Electrons, {\em Phys.\ Rev.\ Lett.} \textbf{19}, 1049  
(1967).

\bibitem{Koester76}
L. Koester:
Verification of the equivalence principle of gravitational and inertial mass  
for the neutron, {\em Phys.\ Rev.} \textbf{D 14}, 907 (1976).

\bibitem{Staudenmannetal80}
J.--L. Staudenmann, S.A. Werner, R. Colella, A.W. Overhauser: Gravity and  
Inertia in Quantum Mechanics, {\it Phys. Rev.} {\bf A 21} 1419 (1980).

\bibitem{HughesDreverExp}
T.E. Chupp, R.J. Hoara, R.A. Loveman, E.R. Oteiza, J.M. Richardson, and M.E.  
Wagshul: Results of a New Test of Local {L}orentz Invariance: A Search for  
Mass Anisotropy in ${}^{21}\hbox{Ne}$, {\em Phys.\ Rev.\ Lett.} \textbf{63},  
1541 (1989).

\bibitem{Laemmerzahl97} C.\ L\"ammerzahl: Constraints on space--time torsion  
from Hughes--Drever experiments, {\it Phys. Lett.} {\bf A 228}, 223
(1997).

\bibitem{HLMashhoon88} B. Mashhoon: Neutron Interferometry
in a Rotating Frame of Reference, {\it Phys.\ Rev.\ Lett.} {\bf 61},
2639 (1988).

\bibitem{HL_Mashhoon95} B. Mashhoon: On the coupling of
intrinsic spin with the rotation of the earth, {\it Phys.\ Lett.}
{\bf A 198}, 9 (1995).

\bibitem{HL_Venemaetal92} B.J. Venema, P.K. Majumder,
S.K. Lamoreaux, B.R. Heckel, and E.N. Fortson: Search for a Coupling of the  
Earth's Gravitational Field to Nuclear Spins in Atomic Mercury, {\it
Phys.\ Rev.\ Lett.} {\bf 68}, 135 (1992).

\bibitem{VessotLevine80}
R.F.C. Vessot, M.W. Levine, E.M. Mattison, E.L. Blomberg, T.E. Hoffmann,  
G.U. Nystrom, B.F. Farrel, R. Decher, P.B. Eby, C.R. Baughter, J.W. Watts,  
D.L. Teuber, and F.D. Wills: Test of Relativistic Gravitation with a  
Space--Borne Hydrogen Maser, {\em Phys.\ Rev.\ Lett.} \textbf{45}, 2081  
(1980).

\bibitem{Grieseretal94}
R. Grieser, R. Klein, G. Huber, S. Dickopf, I. Klaft, P. Knobloch, P. Merz,  
F. Albrecht, M. Grieser, D. Habs, D. Schwalm, and T. K{\"u}hl:
A test of special relativity with stored lithium ions, {\it Appl.\ Phys.}  
\textbf{B 59}, 127 (1994).

\bibitem{PrestageTjoelkerMaleki95}
J.D. Prestage, R.L. Tjoelker, and L. Maleki: Atomic clocks and variations of  
the fine structure constant, {\it Phys.\ Rev.\ Lett.} \textbf{74}, 3511  
(1995).

\bibitem{Schaefer99}
B.E. Schaefer: Severe limits on variations of the speed of light with  
frequency, {\em Phys.\ Rev.\ Lett.} \textbf{82}, 4964 (1999).

\bibitem{HauganKauffmann95}
M.P. Haugan and T.F. Kauffmann:
A New Test of the {E}instein Equivalence Principle and the Isotropy of Space,
{\em Phys.\ Rev.} \textbf{D 52}, 3168 (1995).

\bibitem{Billetetal99} S.D. Biller {\it et al.}: Limits to quantum gravity  
effects on energy dependence of the speed of light from observations of TeV  
flares in active galaxies, {\it Phys.\ Rev.\ Lett.} {\bf 83}, 2108 (1999).
\end{thebibliography}
\end{document}